# Sample-Align-D: A High Performance Multiple Sequence Alignment System using Phylogenetic Sampling and Domain Decomposition

Fahad Saeed and Ashfaq Khokhar

Email: {fsaeed2, ashfaq}@uic.edu
Department of Electrical and Computer Engineering
University Of Illinois at Chicago
Chicago, IL 60607

**Abstract**

*Multiple Sequence Alignment (MSA) is one of the most computationally intensive tasks in Computational Biology. Existing best known solutions for multiple sequence alignment take several hours (in some cases days) of computation time to align, for example, 2000 homologous sequences of average length 300. Inspired by the Sample Sort approach in parallel processing, in this paper we propose a highly scalable multiprocessor solution for the MSA problem in phylogenetically diverse sequences. Our method employs an intelligent scheme to partition the set of sequences into smaller subsets using k-mer count based similarity index, referred to as k-mer rank. Each subset is then independently aligned in parallel using any sequential approach. Further fine tuning of the local alignments is achieved using constraints derived from a global ancestor of the entire set. The proposed Sample-Align-D Algorithm has been implemented on a cluster of workstations using MPI message passing library. The accuracy of the proposed solution has been tested on standard benchmarks such as PREFAB. The accuracy of the alignment produced by our methods is comparable to that of well known sequential MSA techniques. We were able to align 2000 randomly selected sequences from the Methanosarcina acetivorans genome in less than 10 minutes using Sample-Align-D on a 16 node cluster, compared to over 23 hours on sequential MUSCLE system running on a single cluster node.*

## 1. Introduction

Multiple Sequence Alignment (MSA) is a fundamental problem of great significance in computational biology as it provides vital information related to the evolutionary relationships, identify conserved motifs, and improves secondary and tertiary structure prediction for RNA and proteins. In theory, multiple sequence alignment can be achieved using pair-wise alignment, each pair getting alignment score and then maximizing the sum of all the pair-wise alignment scores. Optimizing this score, however, is NP complete [1] and dynamic programming based solutions have complexity of $O(L^N)$, where $N$ is the number of sequences and $L$ is the length of each sequence, thus making such solutions impractical for large number of sequences.

These accurate optimization methods are also very expensive in terms of memory and time. This is why most multiple sequence alignments techniques rely on heuristic algorithms, most popular being CLUSTALW [27], T-Coffee [28], MUSCLE [3, 12] and ProbCons [29]. These heuristics are usually complex combination of ad-hoc procedures mixed with some elements of dynamic programming, thus the resulting methods do not scale well. These methods yield extremely poor performance for very large number of sequences. For example, CLUSTALW [27] is estimated to take 1 year to align 5000 sequences of average length of 350 [3]. MUSCLE is claimed to be the fastest and some what most accurate multiple alignment tool till to date. It claims to align 5000 synthetic sequences of average length 350 in 7 minutes on a contemporary desktop computer [3]. We also performed experiments with real data sets of *Methanosarcina acetivorans genome sequence,* having 5 million base pairs and is by far the largest known archeal genome [31]. The MUSCLE system takes about 1400 minutes (~23 hrs) to align randomly selected 2000 sequences, with average length of 316. Our projection is that MUSCLE system will take more than 30 days to align the whole genome.

The computation demands of these heuristics make the design of parallel approaches to the MSA problem highly desirable. There have been numerous attempts to parallelize existing sequential methods. CLUSTALW [27] is by far the most often parallelized algorithm [4]. James *et. al.* in [5] parallelized CLUSTALW for PC clusters and distributed/shared memory parallel machines. HT Clustal [6] is parallel solution for heterogeneous Multiple Sequence Alignment and MultiClustal [6] is a parallel version of an optimized CLUSTALW. In these solutions, the first two stages, i.e. pair-wise alignment and guide tree, are parallelized, and the third stage, final alignment, is mostly sequential, thus limiting the amount of he achievable speedup [6]. Different modules of the MUSCLE system have also been parallelized [7]. Other parallelization efforts include parallel multiple sequence alignment with phylogeny search by simulated annealing by Zola et al [8], Multithreading Multiple Sequence Alignment by Chaichoompu et al [9] and Schmollinger et al's parallel version of DIALIGN [10]. Although there seems to be a considerable amount of effort to improve the running times for large number of sequences using parallel computing, it must be noted that almost all the existing solutions have been aimed at parallelizing different modules of a known sequential system.

A few attempts have also been made to cut each sequences into pieces and compute the piece wise alignment over all the sequences to achieve multiple sequence alignment. In [22], each sequence is 'broken' in half, and halves are assigned to different processors. The Smith-Waterman [21] algorithm is applied to these divided sequences. The sequences are aligned using dynamic programming algorithms, and then combined using Combine and Extend techniques. The Combine and Extend methods follow certain models defined to achieve alignment of the combination of sequences. These methods pay little or no attention to the quality of the results obtained. The end results have considerable loss of sensitivity. The constraints in these methods are solely defined by the models used, thus limiting the scope of the methods for wide variety of sequences.

In our previous work [34,35], we have developed sampling based sequential and parallel solutions for the alignment of homologous sequences. The sampling techniques used in [34] were based on the assumption that the underlying sequence set has phylogenetically higher correlation. In this paper we propose a solution for aligning phylogenetically *diverse* set of sequences, therefore referred to as Sample-Align-D. The proposed approach is based on domain decomposition where data domain is distributed among processors and local alignments are performed in each processor. At the end, the summary results of the local alignments are shared among all processor, and final adjustments in the alignments are made based on this global information.

Our method draws its motivation from the SampleSort [13] approach that has been introduced to sort large sequence of numbers on multiprocessor or distributed platforms. The sorting and MSA problems share a common characteristic, i.e., any correct solution requires comparison of each pair of data items. In SampleSort, a small sample ($<< N$) representing the entire data set is chosen over distributed partitions using some sampling technique such as Regular Sampling. The sample is then used to define $p$ buckets, where $p$ is the number of processors in the system. The bucket boundaries are made known to all the processors. Each processor then places its local data items into corresponding buckets. The buckets are then individually sorted to achieve an overall sorted sequence. We use a similar sampling approach to redistribute sequences over all the processors based on k-mer rank such that sequences with similar k-mer rank values are available on a single processor. The sequences with similar k-mer ranks are aligned sequentially at different processors. The final alignment is achieved by computing a global ancestor of the underlying homologous set of sequences and profile aligning each sequence with this ancestor.

The rest of the paper is organized as follows: Section 2 gives the description of our method known as Sample-Align-D, the assumptions, and the complexity analysis of the system. Section 3 gives the details of the communication cost and load balancing aspects of the system. Section 4 describes the performance evaluation in terms of execution time, scalability, and quality of the alignments obtained. Section 5 discusses the conclusions and the future work.

## 2. Sample-Align-D

The objective of our work is to develop a highly scalable distributed multiple sequence alignment method based on well-known sequential techniques such that multiple subsets could be aligned in parallel while still achieving global alignment with respects to the entire set. The proposed Sample-Align-D approach uses an idea derived from the SampleSort technique [13] well known in the area of parallel computation to guide the distribution of sequences among processors using regular sampling. We use k-mer rank defined below to achieve the localization of similar sequences on a single processor.

### k-mer Rank:

A k-mer is a contiguous subsequence of length $k$ and related sequences tend to have more k-mer in common than may be expected by chance [3]. The k-mer distance between any two sequences $x_i$ and $x_j$ is defined as follows:

$$r_{i,j} = \sum_{\tau} \tau \min[nx_i(\tau), nx_j(\tau)] / [\min(|x_i|, |x_j|) - k + 1]$$

Here $\tau$ is a $k$-mer of length $k$, $nx_i(\tau)$ and $nx_j(\tau)$ are the number of times $\tau$ occurs in $x_i$ and $x_j$ respectively, and $|x_i|$ and $|x_j|$ are the sequence lengths. We define average k-mer distance of sequence $x_i$ to all the other sequences as follows:

$$D_i = \frac{1}{N} \sum_{j=1}^{N} r_{i,j}$$

Finally, the k-mer rank of a sequence $x_i$ with respect to all the sequence in the data set is defined as follows [15]:

$$R_i = \log(0.1 + D_i)$$

The k-mer distance can be used to rapidly construct phylogenetic trees. For $N$ unaligned sequences of length $L$, the k-mer rank gives an approximate estimate of the fractional identity that has also been used in CLUSTALW. Our motivation for using this type of similarity indexing is to improve processing speed without the need to align the sequences globally. Edgar RC [15] has shown that k-mer similarities correlate well with fractional identity. In the following we first give a very short and intuitive description of the proposed Sample-Align-D approach along with the algorithm.

The main idea in the Sample-Align-D approach is to collect a relatively small sample of sequences (<< $N$) that is representative of the entire data set of $N$ sequences. The k-mer rank is computed for all the sequences in the sample and then this rank is used to redistribute and locally align the datasets in each processor in parallel using a sequential MSA algorithm. Further fine tuning of the alignment is performed using a global ancestor template and thus achieving global alignment. The global ancestor is computed using local ancestors that are available at the end of the local alignment phase with in each processor. Based on this approach, the complexity of the alignment is reduced to aligning $p$ sets of sequences, each of size $N/p$ with some, additional cost incurred on communication and fine tuning. A more detailed algorithm is outlined below.

---

**Sample-Align-D (Sequences N)**
$p$ processor are being used for computation

    **Input** ← N sequences of amino acids x1, x2 … xN:
    **Output** ← Gaps are inserted in each of the x's so that
- **All sequences have the same length**
- **Score of the global map is maximum**

    ➢ Assume *N/p* sequences on each of the *p* processors
    ➢ Locally compute k-mer rank of all the sequences in each processor

- Sort the sequences locally in each processor based on k-mer rank
- Choose a sample set of *k* sequences in each processor, where $k << N/p$
- Send the *k* samples from each processor to all the processors.
- Compute the k-mer rank of each sequence against the $k*p$ samples.
- Sort the sequences locally in each processor based on the new k-mer rank
- Using regular sampling, choose p-1 sequences from each processor and send only their ranks to a root processor.
- Sort all the $p*(p-1)$ ranks at the root processors and divide the range of ranks into *p* buckets.
- Send the bucket boundaries to all the processors.
- Redistributed sequences among processors such that sequences with k-mer rank in the range of bucket *i* are accumulated at processor *i*, where $0 > i < p+1$.
- Align sequences in each processor using any sequential multiple alignment system
- Broadcast the Local Ancestor to the root processor
- Determine 'Global' Ancestor GA at the root processor by aligning 'local' ancestors received from all the processors
- Broadcast GA to all the processors
- Realign each of the sequences in *p* processors based on ancestor GA using profile-profile alignment .i.e. Each of the profiles of aligned sequences are tweaked using the ancestor profile, with constraints.
- Glue all the aligned sequences at the root processor.
- END

## 2.1 Assumptions and Limitations

We state certain assumptions and limitations that are typical of parallel systems/ alignment tools and are not overly restrictive or extensive.

- Currently, the method is only being tested for homologous sequences. The method is also envisioned to work best when the sequences are not highly divergent.
- It is assumed that the sequences similarity is uniformly distributed and redistribution step in the above algorithm provides statistical guarantee of uniform load over all the processors.
- The method would work best when the number of sequences is large, so that the inter-processor communication is much less than the time required to align the set of sequences on a single processor.

## 2.2 The Sample-Align-D Algorithm: Details

It is assumed that the input consists of *N* sequences of amino acids $x_1, x_2... x_N$. The output of the algorithm is again a set of *N* sequences such that the gaps are inserted in a way that the sequences have equal length and score of the global map is maximized. The score is calculated as the sum of pairs' scores. The *N* sequences are divided equally among *p* processors, i.e. each processor is assigned *N/p* sequences. A similarity index based on k-mer rank of each sequence is computed locally on each processor.

### 2.3.1. Globalised K-mer Rank

In our earlier paper [34], we assumed phylogenetically higher correlation among sequences. Therefore, k-mer rank computed in each processor locally, independent of the sequences at the other processors is similar to the rank that might have been determined globally considering all the sequences. We showed in [34] that this is true when the sequences are not highly divergent.

If the sequences are not highly divergent, k-mer rank computed using this distributed approach may be very different compared to the centralized case where rank of each sequence is computed by considering all the *N* sequences.

In order to address this diversity, we collect *k* sample sequences from each processor such that these *k* samples represent the corresponding set of *N/p* sequences, yielding a total of $k*p$ samples. Collectively, it is safe to assume that these $k*p$ samples represent the entire set of *N* sequences. We use these $k*p$ sequences to built a phylogenetic tree, and for each sequence compute its k-mer rank using this tree. Thus the rank computation for each sequence is against a global sample.

Subsequently, redistribution based on this sampling technique also ensures that sequences accumulated in each processor are 'similar' to each other. In Fig. 1 we compare the rank computed using samples only (referred to as globalized rank) with the rank computed using all the sequences (referred to as centralized rank).

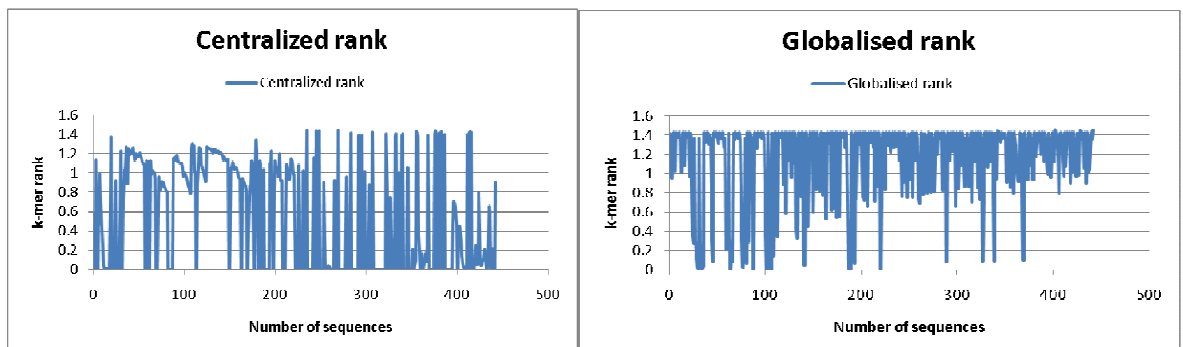

**Fig. 1.** Distribution of k-mer ranks for 500 sequences when computed on a central system and globalised kmer systems.

The statistics of the two approaches for 5000 sequences are presented in Table 1. As can be seen that the standard deviation for the two sets of ranks is 0.58 and that the average of the globalised approach is higher than that of the centralized approach. This is due to the fact that in the globalized approach each sequence is compared against a small set of sequences, where as in the centralized approach each sequences is compared against all the $N$ sequences, yielding larger variations in the kmer rank and making the average smaller.

| (Maximum, Minimum) Central | ( 1.44827    ,0.0) |
|---|---|
| Average Centralized | 0.722962 |
| (Maximum, Minimum) Globalized | (1.46207,0.0) |
| Average Globalized | 1.11302 |
| Variance w.r.t. Centralized | 0.33190 |
| Standard Dev. w.r.t Centralized | 0.576377 |

**Table 1.** Statistical comparison of the k-mer rank computed on a Gloablized system vs Central system.

### 2.3.2 Redistribution Based on k-mer Rank

Each processor sorts its $w=N/p$ sequences based on k-mer ranks using a sequential sorting algorithm. From each of the $p$ locally sorted lists, $k = (p-1)$ evenly spaced samples are chosen. The k-mer ranks of these $p-1$ samples (pivots) divide the local set of sequences into $p$ ordered subsets. The k-mer ranks of these $p-1$ samples from all the processors are gathered at the root processor yielding a set Y of size $p(p-1)$. This regular sampled set Y is sorted to compute the ordered list $Y_1, Y_2, Y_3 … Y_{p(p-1)}$ determining the range of k-mer ranks over all the processors. Then $Y_{p/2}, Y_{p+p/2} … Y_{(p-2)p+p/2}$ are chosen as pivots ($p$ in total) dividing the k-mer rank range into $p$ buckets. These pivots are then broadcast to all the processors. Each processor sends the sequences having k-mer rank in the range of bucket $i$ to processor $i$. For the bound on the size of the dataset in each processor after redistribution, we refer to the analysis in Section 3.

### 2.3.3 The Alignment

Next a sequential MSA program is run on each processor. Since our ultimate goal is to have a sequence alignment for $N$ sequences and not on some subset of N sequences. Therefore, a way has to be defined that would concatenate these 'chunks' of aligned sequences so that a 'global' alignment of multiple sequences can be obtained. In [12] it has been observed that multiple sequence alignment for homologous sequences can be obtained by aligning each sequence to the root profile. This approach is similar to the one used in the PSI-BLAST [19], where a profile is used to align any query sequence with the sequences that have generated the profile. A similar idea for ancestor constrained multiple sequence alignment has been proposed, although without domain decomposition, for progressive alignment [33].

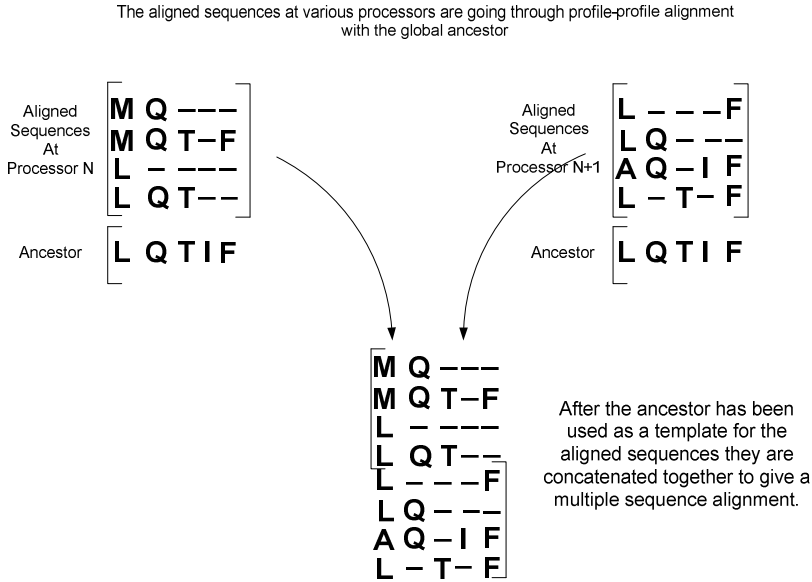

**Fig. 2.** Example of Ancestor profile being used to tweak locally aligned profiles in different cluster nodes.

We use a similar concept along with domain decomposition of the sequences. We extract the local ancestor from each processor after locally aligning each subset in parallel. All of these local ancestors are collected at the root processor and are aligned using a sequential multiple sequence alignment algorithm. The ancestor of all the local ancestors, referred to as the 'global' ancestor, is then broadcast back to all the processors. The 'global' ancestor is then used to perform a profile-profile alignment using the method in [12], i.e. each of the locally aligned sequences (referred to as profile) in the processor is aligned with the global ancestor profile. The profile-profile alignment with the template of the ancestor is performed, to get a better SP score and hence a multiple sequence alignment. The kind of fine tuning that may be expected from the ancestor can be depicted in Fig. 2.

As shown in Fig. 2, there are two sets of sequences that are aligned independent of each other. To perform a one global alignment of these multiple sequences, these subsets of independently aligned sequences are tweaked, using the global ancestor as a template. After this step, the tweaked sequences are just 'joined' together and SP score is obtained by just concatenating the sequences.

## 3. Analysis of Computation and Communication Costs:

For analysis purposes we assume that the sequential MSA algorithm being used is the MUSCLE system [12]. Therefore, the complexity computations of sequential components are based on the analysis given in [12]. Here the assumption is that initially each processor has $w = N/p$ sequences, where $N$ is the number of sequences and $p$ is the number of processors. The average length of a sequence is $L$. In the following we first out line all the computation costs and storage requirement. This is followed by the analysis of the communication overhead.

**Computation Costs:**

| STEP | O (Time) | O (Space) |
|---|---|---|
| 1. k-mer rank computation on ($w=N/p$) sequences | $w^2 L$ | $w+L$ |
| 2. Sorting of $N/p$ sequences based on k-mer rank | $w\ log w$ | $log\ w$ |
| 3. Sample $k = p-1$ sequences | $w$ | $p$ |
| 4. k-mer rank computation of ($k*p$) sequences | $p^4 L$ | $p^2+L$ |

in root processor

| | | |
|---|---|---|
| 5. Sorting of $k*p$ sample k-mer ranks | $(k*p)\log(k*p)$ | $\log(k*p)$ |
| 6. K-mer rank computation of each of (w=N/p) Sequences against k*p samples | $w[(k*p+1)^2 L]$ | $w(k*p+L)$ |
| 7. MUSCLE executed on ($w=N/p$) sequences in parallel. | $w^4+wL^2$ | $w^2+L^2$ |
| 8. Ancestor extraction from each of the $p$ processors + export to the root processor. | $p^2$ | $p^2$ |
| 9. MUSCLE executed on local ancestors ($p$ elements) | $(p)^4+(p)L^2$ | $(p)^2 + L^2$ |
| 10. Profile alignment with all combined aligned sequences on each of the processor | $wL^2$ | $w$ |
| TOTAL Computation Cost (for $w = N/p$) | $O((N/p)^4+ (N/p) L^2)$ | $O((N/p)^2+ L^2)$ |

**Communication Costs:**

No matter how powerful a machine may be, inter-processor communication overhead is a factor that limits the performance of a distributed message passing parallel systems [24]. Fortunately, the communication cost of our system is much less than the cost of the alignments. Essentially, the proposed Sample-Align-D algorithm has two rounds of communication. . In the first round, samples are collected at the root processor and pivots broadcast from the root processor. In the second round, sequences are redistributed to achieve better alignments and balanced load distribution. For the analysis of the communication costs we have adopted the coarse grained computation model assumed in [20, 16, 2]. However, we ignore the message start up costs and assume unit time to transmit each data byte.

We have assumed Regular Sampling strategy due to following reasons:

1. The strategy is independent of the distribution of original data, compared to some other strategies such as Huang and Chow [25].
2. It helps in partitioning of data into ordered subsets of approx. equal size. This presents an efficient strategy for load balancing as unequal number of sequences on different processors would mean unequal computation load, leading to poor performance. In the presence of data skew, regular sampling guarantees that no processor computes more than ($2N/p$) sequences [26].
3. It has been shown in [26] that regular sampling yields optimal partitioning results as long $N>p^3$, i.e., the number of data items $N$ is much larger than the number of processors $p$, which would be a normal case in the MSA application.

*First Communication Round:*
Assuming $k = p-1$, i.e., each processor chooses $p-1$ samples, the complexity of the first phase is $O(p^2L)+ O(p\log p)+ O(k*p\log p)$, where $O(p^2L)$ is the time to collect $p(p-1)$ samples of size $L$ each at the root processors, $O(p\log p)$ is the time required to broadcast $p-1$ pivots to all the processor and ($k*p\log p$) is the time required to broadcast k*p sequences to all the processors.

*Second Communication Round:*
In the second round each processor sends the sequences having k-mer rank in the range of bucket *i* to processor *i*. Each processor partition its blocks into p sub-blocks, one for each processor, using pivots as bucket boundaries. Each processor then sends the sub-blocks to the appropriate processor. These sub-blocks can vary from 0 to *N/p* sequences depending on the initial data distribution. Taking the average case where the elements in the processor are distributed uniformly, then each sub-block size is $N/p^2$. Thus this step would require $O(N/p)$ time assuming an all-to-all personalized broadcast communication primitive [16, 2]. However, in the following we show that based on regular sampling no processor will receive more than $2N/p$ elements in the worst case. Therefore still the overall communication cost will be O ($N/pL$).

Let's denote the pivots chosen in the first phase by the array: y1, y2, y3...yp-1. Consider processor i = 1, *where* $1 \leq i \leq p$, all the data to be aligned by processor 1 must be $\leq y_1$ in terms of it k-mer rank. Since there are $p^2$-p-p/2 sequences in the sample that have k-mer rank > $y_1$, correspondingly there are at least $(p^2$-p-p/2) w/p sequences in the entire data set whose rank is > $y_1$. In other words there are $N$-$(p^2$-p-p/2) w/p= (p+p/2) w/p < 2w sequences in the datasets which have k-mer rank $\leq y_1$. The size of data to be locally aligned by any processor is therefore always less than $2w$. Due to page limitations, we refer to [26] for further details on the analysis of this bound. The collection of the *p* local ancestors at the root processor and the broadcast of the global ancestor will cost O ($Llogp$) communication overhead each. Therefore the total communication cost is: O $(p^2L)$ + O $(plogp)$ + O $(N/pL)$ + O ($Llogp$)

The total asymptotic time complexity T of the algorithm would be:
= $\underline{O(N/p)^4 + O(N/p) L^2}$ + $\underline{O(p^2L) + O(plogp + O(N/pL) + O(Llog p) + O(k*plogp)}$
      Computation cost                        Communication cost

= $O((N/p)^4 + (N/p) L^2 + (p^2L) + (N/pL))$

## 4. Performance Evaluation

The performance evaluation of the Sample-Align-D Algorithm is carried out on a Beowulf Cluster consisting of 16 Pentium III processors, each running at 550 MHz, with 2 levels of cache (L1: 16K and L2: 512K), and 384 MB DRAM memory. As for the interconnection network, the system uses Intel Gigabit NIC's on each cluster node. The operating system on each node is RedHat Linux 7.3 (Kernel level: 2.4.20-28.7). For performance evaluation we have used both synthetic and real data sets.

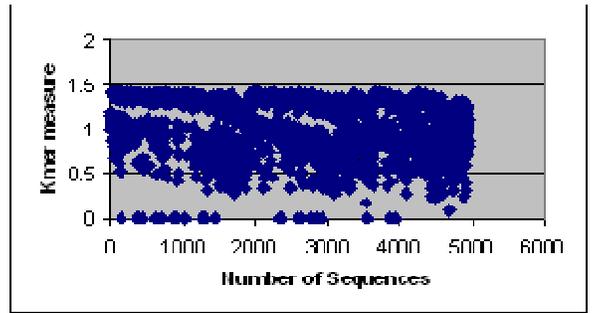

**Fig. 3.** Distribution of k-mer rank of the sequences used in the experiments

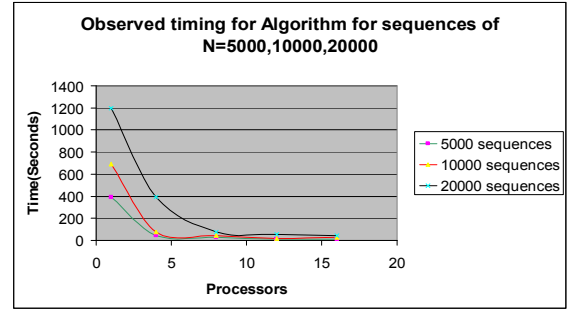

**Fig. 4.** Scalability of the execution time with respect to the number of processors being used.

To investigate the resource requirements and the execution time on different size inputs we have used the synthetic data set generated using rose sequence generator [14]. Three sets of sequences (N=5000, 10000, and 20000) were generated using the standard input parameters for the rose generator. The average sequence length was set to be 300 and the relatedness was set to be 800. This relatedness value assured that the sequences thus generated were in fact not very close to each other and may resemble the real dataset of protein sequences. Furthermore, in these experiments it was made sure that the k-mer rank distribution for the sequences is in general evenly distributed. A sample of k-mer rank distribution for N = 5000 sequences used in our experiments is shown in Fig. 3.

After the sequences were generated according to the experimental setup, the files were divided into equal parts and 'placed' on the cluster nodes' hard drives prior to the experiments. Then an instance of Sample-Align-D Algorithm was initiated on each

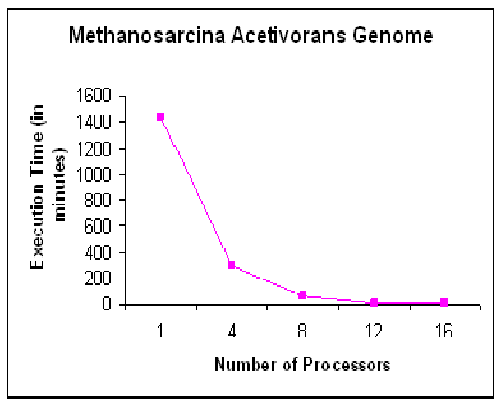

**Fig. 6.** Execution time on randomly selected 2000 sequences from the *Methanosarcina Acetivorans genome*.

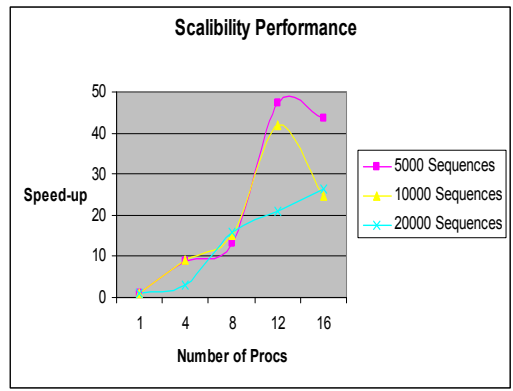

**Fig. 5.** Super linear speed-up for Sample Align-D with increasing number of of the nodes and the time required for the actual alignment was noted. We were able to align 20000 sequences in just around 25 seconds. There are no reports of aligning this huge number of sequences in the literature to the best of authors' knowledge. The best that the authors found was for N=5000 sequences [12]. T-coffee [28] is reported to not able to handle more than $10^2$ sequences. MAFFT [23] script FFTNSI is reported to align 5000 sequences in 10 minutes and MUSCLE itself without refinement in 7minutes. It is anticipated that with refinement included, MUSCLE is bound to take the same amount as FFTNSI. It is also estimated that CLUSTALW [27] would take approximately 1 year to align these many sequences [12]. As shown in Fig. 4, in the case of Sample–Align, the execution time decreases sharply with the increase in the number of processors. We got super linear speed-up for the Sample-Align-D and the observed speedup curves are shown in Fig. 5. This is primarily because the computation complexity decreases by O ($p^4$) with the increase in number of processor. It can be observed, however, that for the datasets of N=5000 and 10000, the speedup curve goes up for 4, 8 and 12 processors but deteriorates when all the 16 processors are used. The slowdowns indicate that the granularity of work assigned to each processor decreases. With the increase of the dataset to 20000, we get much better speedup curves.

We have also experimented with real protein sequences from the *Methanosarcina acetivorans genome.* Fig. 6 depicts the execution time of aligning randomly selected 2000 sequences from the *Methanosarcina acetivorans genome* using different number of processors. . Note that it took more than 23 hours to align this set of sequences using the sequential MUSCLE system on a single cluster node, whereas it took only 9.82 minutes to align the same number of sequences using the proposed Sample-Align-D algorithm. This is a 142 fold speedup using 16 processors!

### 4.1. Quality Assessment

We have used the PREFAB benchmark to assess the quality of the alignments produced by Sample-Align-D Algorithm We have used the accuracy measure Q [3], defined as, the number of correctly aligned residue pairs divided by the number of residue pairs in the reference alignment. The results from this benchmark are presented below in Table 2. For the Sample-Algorithm, results correspond to execution on a 4 processor cluster system.

| METHOD | Q-Score |
|---|---|
| **Sample-Align-D** | 0.544 |
| **MUSCLE** | 0.645 |
| **MUSCLE-p** | 0.634 |
| **T-Coffee** | 0.615 |
| **NWNSI** | 0.615 |
| **FFTNSI** | 0.591 |
| **CLUSTALW** | 0.563 |

**Table 2.** Q Scores obtained for each method using PREFAB.

As can be seen from the values above that our method provides quality of alignment comparable to the other well-known methods. For PREFAB, Sample-Align-D Algorithm yields quality very close to that of CLUSTALW.

It should be noted that we are getting quality[1] comparable to CLUSTALW and execution time better than MUSCLE itself. We believe that quality is much higher in our case when the sequences are large in number. However, due to the absence of large size benchmark datasets we cannot report supporting results at the time of this publication. In the case of PREFAB, it contains 1000 set of approx. 20-30 sequences each. Each set is aligned independently of the other sets to access the quality of the alignment program on multiple set of sequences of varying divergence. In the case of the proposed Sample-Align-D Algorithm, partitioning each set of 20 to 30 sequences even on a 4 processor system is too fine grain to access the true quality of alignment. More detailed quality analysis results will be presented in the full version of this paper. A snap shot of the alignment produced by the Sample-Align-D Algorithm for the genome sequences is given in Fig. 7.

## 5  Conclusions and Future Research

We have addressed the long standing problem of aligning large number of multiple sequences in a reasonable amount of time. We have addressed the problem using a completely distributed approach to the problem, using high scalable techniques similar to sample sort. The sequences are distributed among the processors according to k-mer rank and are aligned in a distributed manner independently of the other sequences. The independently aligned sequences are then aligned with the global ancestor as is described in the paper. The sequences are then joined in a root processor giving a meaningful alignment. Our results show super linear speed-up with comparable quality of alignment.

Currently we are working on accessing the quality of the method using other standard benchmarks such as BAliBASE, SMART and SABmark. It must be noted however, that these benchmarks are not designed to access the quality of the alignments produced in a distributed manner and the size of these benchmarks may limit the accuracy of the quality accessed. Therefore, it would be desirable to develop benchmarks that may be used to access the quality of these alignments formulated using distributed systems as ours. We are working on sequential heuristics to improve the quality of the alignment produced from the methods discussed above. These heuristics may then be further parallelized to incorporate in the distributed approach presented in this paper.

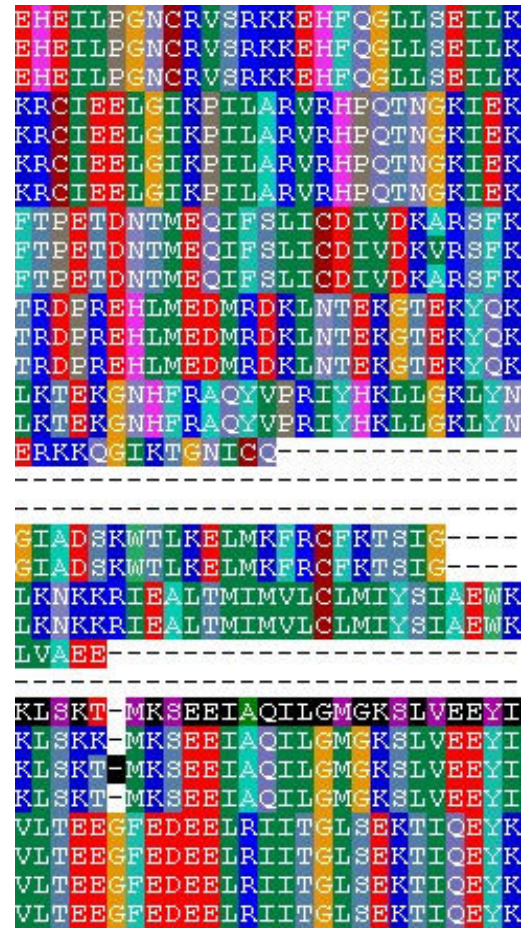

**Fig. 7.** A Snap shot of the alignment produced by Sample-Align Algorithm for the sequences in the *Methanosarcina acetivorans genome*

Another area that is interesting to explore is the kind of sequences and families that may be aligned using the methods discussed, with reasonable accuracy. It can be seen however that there might always be a need to refine the 'global' multiple sequence alignment for some of the most divergent families and sequences. An efficient method to do that with small time complexity would be necessary for some families of sequences being aligned, thus making the system working for a wide range of families but still keeping the advantage of high scalability and performance.

---

[1] Some of the sequence scores were discarded in the automatic quality estimation process.